\documentstyle[12pt]{article}
\if@twoside\oddsidemargin25mm
\else\oddsidemargin25mm\evensidemargin25mm\marginparwidth25mm\fi
\def\thefootnote{\fnsymbol{footnote}}
\def\bl{\Big\{}
\def\br{\Big\}}
\def\bpl{\Big(}
\def\bpr{\Big)}
\def\ve{\varepsilon}
\def\t{\theta}
\def\vphi{\varphi}

\def\O{\Omega}
\def\F{{\cal F}}

\def\der{\partial}
\def\bq{\begin{equation}}
\def\eq{\end{equation}}
\def\brr{\begin{eqnarray}}
\def\err{\end{eqnarray}}
\def\ba{\left(\begin{array}}
\def\ea{\end{array}\right)}

\def\Dslash{\hbox{\ooalign{$\displaystyle D$\cr$\hspace{.03in}/$}}}

\def\Eslash{\hbox{\ooalign{$\displaystyle E$\cr$\hspace{.03in}/$}}}

\def\Vslash{\hbox{\ooalign{$\displaystyle V$\cr$\hspace{.02in}/$}}}
\begin{document}
\newcommand{\dr}{\raise.3ex\hbox{$\stackrel{\leftarrow}{\partial }$}{}}
\newcommand{\dl}{\raise.3ex\hbox{$\stackrel{\rightarrow}{\partial}$}{}}
\newcommand{\eqn}[1]{(\ref{#1})}
\newcommand{\ft}[2]{{\textstyle\frac{#1}{#2}}}
\newcommand{\dkt}{\delta _{KT}}
\newcommand{\QED}{{\hspace*{\fill}\rule{2mm}{2mm}\linebreak}}
\renewcommand{\theequation}{\thesection.\arabic{equation}}
\csname @addtoreset\endcsname{equation}{section}
\newsavebox{\uuunit}
\sbox{\uuunit}
    {\setlength{\unitlength}{0.825em}
     \begin{picture}(0.6,0.7)
        \thinlines
        \put(0,0){\line(1,0){0.5}}
        \put(0.15,0){\line(0,1){0.7}}
        \put(0.35,0){\line(0,1){0.8}}
       \multiput(0.3,0.8)(-0.04,-0.02){12}{\rule{0.5pt}{0.5pt}}
     \end {picture}}
\newcommand {\unity}{\mathord{\!\usebox{\uuunit}}}
\newcommand  {\Rbar} {{\mbox{\rm$\mbox{I}\!\mbox{R}$}}}
\newcommand  {\Hbar} {{\mbox{\rm$\mbox{I}\!\mbox{H}$}}}
\newcommand {\Cbar}
    {\mathord{\setlength{\unitlength}{1em}
     \begin{picture}(0.6,0.7)(-0.1,0)
        \put(-0.1,0){\rm C}
        \thicklines
        \put(0.2,0.05){\line(0,1){0.55}}
     \end {picture}}}
\newsavebox{\zzzbar}
\sbox{\zzzbar}
  {\setlength{\unitlength}{0.9em}
  \begin{picture}(0.6,0.7)
  \thinlines
  \put(0,0){\line(1,0){0.6}}
  \put(0,0.75){\line(1,0){0.575}}
  \multiput(0,0)(0.0125,0.025){30}{\rule{0.3pt}{0.3pt}}
  \multiput(0.2,0)(0.0125,0.025){30}{\rule{0.3pt}{0.3pt}}
  \put(0,0.75){\line(0,-1){0.15}}
  \put(0.015,0.75){\line(0,-1){0.1}}
  \put(0.03,0.75){\line(0,-1){0.075}}
  \put(0.045,0.75){\line(0,-1){0.05}}
  \put(0.05,0.75){\line(0,-1){0.025}}
  \put(0.6,0){\line(0,1){0.15}}
  \put(0.585,0){\line(0,1){0.1}}
  \put(0.57,0){\line(0,1){0.075}}
  \put(0.555,0){\line(0,1){0.05}}
  \put(0.55,0){\line(0,1){0.025}}
  \end{picture}}
\newcommand{\Zbar}{\mathord{\!{\usebox{\zzzbar}}}}
\newcommand{\Ka}{K\"ahler}
\newcommand{\qu}{quaternionic}
\def\ib{{\bar \imath}}
\def\jb{{\bar \jmath}}
\renewcommand{\sp}{{Sp\left( 2n+2,\Rbar\right)}}
\renewcommand{\a}{\alpha}
\renewcommand{\b}{\beta}
\renewcommand{\c}{\gamma}
\renewcommand{\d}{\delta}
\newcommand{\pa}{\partial}
\newcommand{\g}{\gamma}
\newcommand{\G}{\Gamma}
\newcommand{\A}{\Alpha}
\newcommand{\B}{\Beta}
\newcommand{\D}{\Delta}
\newcommand{\e}{\epsilon}
\newcommand{\E}{\Epsilon}
\newcommand{\z}{\zeta}
\newcommand{\Z}{\Zeta}
\newcommand{\K}{\Kappa}
\renewcommand{\l}{\lambda}
\renewcommand{\L}{\Lambda}
\newcommand{\La}{\Lambda}
\newcommand{\m}{\mu}
\newcommand{\M}{\Mu}
\newcommand{\n}{\nu}
\newcommand{\N}{\Nu}
\newcommand{\x}{\chi}
\newcommand{\X}{\Chi}
\newcommand{\p}{\pi}
\newcommand{\R}{\Rho}
\newcommand{\s}{\sigma}
\renewcommand{\S}{\Sigma}
\newcommand{\T}{\Tau}
\newcommand{\y}{\upsilon}
\newcommand{\Y}{\upsilon}
\renewcommand{\o}{\omega}
\newcommand{\q}{\theta}
\newcommand{\h}{\eta}
\begin{titlepage}
\begin{flushright} KUL-TF-96/16\\ THU-96/41\\
hep-th/9612203
\end{flushright}
\vfill
\begin{center}{\LARGE\bf Chern-Simons Couplings
and Inequivalent \\[1.5mm]
Vector-Tensor Multiplets} \\
\vfill
{\large P. Claus\footnote{Wetenschappelijk Medewerker, NFWO, Belgium}}
{,}\,\,
{\large P. Termonia}\\[1.5mm]
{\small
     Instituut voor Theoretische Fysica - Katholieke Universiteit Leuven\\
     Celestijnenlaan 200D B--3001 Leuven, Belgium}\\[1.4cm]
{\large B. de Wit}
\,\,{and}\,\,
{\large M. Faux} \\[1.5mm]
{\small Institute for Theoretical Physics - Utrecht University\\
     Princetonplein 5, 3508 TA Utrecht, The Netherlands} \vspace{7mm}
\end{center}
\vfill
\begin{center} {\bf Abstract}\end{center}
{\small
The off-shell vector-tensor multiplet is considered in an arbitrary background of
$N=2$ vector supermultiplets.  We establish the existence of two
inequivalent versions, characterized by different Chern-Simons
couplings. In one version the vector field of the vector-tensor
multiplet is contained quadratically in the Chern-Simons term, which
implies nonlinear terms in the supersymmetry transformations and
equations of motion.  In the second version, which requires a
background of at least two abelian vector supermultiplets, the supersymmetry
transformations remain at most linear in the vector-tensor components. This
version is of the type known to arise from reduction of
tensor supermultiplets in six dimensions. Our work applies to any number
of vector-tensor multiplets. }
\vspace{7mm}
\flushleft{December 1996}
\end{titlepage}
\renewcommand\thefootnote{\arabic{footnote}}
\setcounter{footnote}{0}
\noindent
\section{Introduction}
Supergravity actions are important tools in the study of string
compactifications. Although physically relevant compactifications
have a lower degree of symmetry, the more restrictive environment
of $N=2$ supersymmetry provides a rich
testing ground for exploring
perturbative and nonperturbative features of string theory,
such as various kinds of
dualities.  An important role is played by the existence of
different matter supermultiplets, whose mutual interactions
are subject to stringent constraints.  For vector and hyper multiplets
these constraints form the basis for specific nonrenormalization theorems.
Multiplets with two-form gauge potentials exhibit even more
restrictive interactions, and presumably stronger nonrenormalization
theorems.  Such multiplets occur universally
in string theory, as the world-sheet formulation naturally incorporates a
target-space tensor gauge field.  In four-dimensional effective field
theories, this tensor field is commonly represented by a (pseudo)scalar
``axion" field, which is obtained by performing a duality transformation
on the tensor.  The extra restrictions are then manifest
in terms of an inevitible Peccei-Quinn invariance, which leads to
specially constrained sigma-model geometries. In the presence of
vector gauge fields, tensor gauge fields tend to
couple to Chern-Simons forms.

In heterotic $N=2$ four-dimensional supersymmetric string vacua,
the axion/dilaton complex resides in a vector-tensor multiplet
\cite{sohnius,DWKLL}. Off shell, this multiplet comprises a real
scalar, a vector gauge field, a tensor gauge field, a real
auxiliary scalar and a doublet of Majorana spinors.
At the linearized level it can be obtained from
reduction of a tensor supermultiplet in six spacetime dimensions.
With respect to $N=1$ supersymmetry, a vector-tensor multiplet
decomposes into a tensor and a vector multiplet.
The scalar and tensor components correspond to the dilaton and
axion, which, after a duality conversion, are combined into a
complex scalar dilaton field belonging to a vector
supermultiplet. As the couplings of vector multiplets to supergravity
have been extensively explored, it is convenient to work in
such a dual formulation.
However, it should be kept in mind that the couplings
are much less restrictive in this context, and may not fully
capture the relevant restrictions of the vector-tensor formulation.
Thus, in this paper, we are motivated by the characterization of
the heterotic axion/dilaton system in terms of an
$N=2$ effective action, within the more restrictive, and presumably more
appropriate, context of the vector-tensor multiplet.
Our results should be regarded as an extension of the work reported in
\cite{vt1}. Rather surprisingly, but in line with the result of
[3], we encounter couplings that are inconsistent with the
typical couplings of the heterotic axion/dilaton complex. The couplings
that we find appear in two varieties and instead play a role
for the nonperturbative corrections to the effective Lagrangian for
heterotic $N=2$ supersymmetric string compactifications
\cite{LSTY}.
One type of coupling could be of six-dimensional
origin \cite{FerMinSag}. For the relevance of nonperturbative
phases with extra six-dimensional tensor multiplets,
see \cite{LSTY, FerMinSag} and references quoted therein.

An off-shell multiplet based on $8+8$ degrees of freedom,
which includes one tensor and one vector gauge field, must
have an (off-shell) central charge.  Alternative off-shell
formulations without a central charge require more,
possibly infinite, degrees of freedom.
Harmonic superspace would provide a natural setting
for describing the latter case. In this paper we discuss
the formulation with a central charge,
which follows a similar pattern as was found for hypermultiplets
\cite{DWLVP}.
Specifically, the central charge acts as a translation
operator, which links an infinite hierarchy of essentially
identical multiplets.  A system of constraints then renders
these additional multiplets dependent, in such a way that
retains precisely $8+8$ off-shell degrees of freedom.
The linearized version of this system has
been discussed in a superspace context in \cite{HOW}.

Some comparison can be made between our results and previous
results pertaining to hypermultiplets \cite{DWLVP}.
There exist quaternionic geometries known to
hypermultiplet theories, which are inaccessible by
the techniques which we employ. These restrictions have never been
fully understood. In principle these could also occur for
vector-tensor multiplets. As we will discuss in due course
this seems indeed the case. Unlike in the hypermultiplet case,
however, our approach yields the more complicated variants of the
vector-tensor multiplet couplings, whereas some qualitatively simple
ones do not fit into our framework. Here we have in mind certain
theories that can be found (on-shell) by dimensional reduction
of five-dimensional Einstein-Maxwell supergravity, after
converting, in five dimensions, one of the vector fields into a
tensor gauge field. On the other hand, some of our couplings are
definitely not obtainable from higher-dimensional theories.  Clearly the
presentation in this paper constitutes only a first word on this
issue and we feel that a further elucidation of these
restrictions is highly desirable.

An important observation is that local supersymmetry requires a
central charge to be realized locally.
This necessitates a background of at least a single vector multiplet
to provide the gauge field for the central charge transformation.
In \cite{vt1} the vector-tensor multiplet was
considered in a background of a single vector multiplet with a
local off-shell central charge.
This resulted in the intricate hierarchical structure described above.
Closure of the combined gauge and supersymmetry algebra then required a
Chern-Simons coupling between the tensor field and the vector field of the vector-tensor multiplet itself.
This in turn induced unavoidable nonlinearities into the
supersymmetry transformation rules and into the action, in terms
of the vector-tensor components.

In this paper we generalize the work of \cite{vt1}
by introducing a more general Chern-Simons term in an extended
background of several vector multiplets.  Crucial differences
then occur. The most conspicuous is that
in a background consisting of at least two vector multiplets,
there exist two inequivalent classes of vector-tensor multiplets.
In the first class the transformation rules are nonlinear.
The theory considered in \cite{vt1} belongs to this class.
In the other class the transformation rules remain at most linear
in the vector-tensor components so that the action is quadratic.
The distinction between these cases is encoded in the particular Chern-Simons
coupling chosen for the tensor field. In this way we
are able to treat both the nonlinear and the linear versions in a
common framework, and establish that, in truth,
there are two inequivalent vector-tensor multiplets in four dimensions
with $8+8$ off-shell degrees of freedom.

In our approach, we also exploit the presence of the vector multiplets
for another purpose. We require all couplings to be invariant under constant
scale and $U(1)$ transformations, so as to facilitate the coupling to
supergravity via the superconformal multiplet calculus.  Imposing
these additional symmetries does not represent a significant
restriction, as we can always freeze some of the
background vector multiplets to a constant, a procedure that
preserves supersymmetry and induces a breakdown of scale and
chiral symmetry. The supergravity
couplings will appear in a separate publication \cite{vtsg},
together with a more detailed analysis of the couplings that seem
to be outside the present framework.

In section~2 of this paper we exhibit the central-charge hierarchy
and features of the  Chern-Simons couplings. In section~3 we give
the supersymmetry transformations for the two classes of
vector-tensor multiplets. In section 4 we present the construction
of a linear multiplet from the vector-tensor multiplet, which enables
the construction of a supersymmetric action, which is presented in
section 5, where we also give the dual formulations of the
corresponding theories in terms of vector supermultiplets.

\setcounter{equation}{0}
\section{Central charge, gauge structure and Chern-Simons couplings}
The vector-tensor multiplet comprises a scalar field $\phi$,
a vector gauge field $V_\mu$, a tensor gauge field $B_{\mu\nu}$,
a doublet of spinors $\l_i$, and an auxiliary scalar field.
As explained in the introduction, since this description
involves $8+8$ off-shell degrees of freedom, it
must incorporate an off-shell central charge. Infinitesimally,
this charge acts as $\d_{{\rm z}}\,\phi=z\phi^{({\rm z})}$.
Successive applications then generate a sequence of translations,
\bq \phi\longrightarrow\phi^{({\rm z})}\longrightarrow
\phi^{({\rm zz})} \longrightarrow {\rm etc}\,,
\label{hierarchy}\eq
and similarly on all other fields\footnote{%
     A hierarchy such as (\ref{hierarchy})
     arises naturally when starting from a five-dimensional
     supersymmetric theory with one compactified coordinate,
     but this interpretation is not essential.}.
It turns out that $\phi^{({\rm z})}$ corresponds to the auxiliary field.
All other objects in the hierarchy, $\phi^{({\rm zz})}$,
$V_{\mu}^{({\rm z})}$, $V_\mu^{({\rm zz})}$, etcetera,
are dependent, and are given by particular combinations of the
independent fields.  This is enforced by a set of constraints, which
we exhibit below.  The central charge then acts so as to generate a
sequence of multiplets, which are not independent; there are no new
degrees of freedom beyond the $8+8$ described previously.

In addition to the central charge, the vector-tensor multiplet is
subject to a pair of gauge transformations, consisting of
a tensor transformation, with parameter $\Lambda_\mu$,
under which $B_{\mu\nu}\rightarrow
B_{\mu\nu}+\der_{[\mu}\Lambda_{\nu]}$,
and a vector transformation, with parameter $\t^1$,
under which $V_\mu\rightarrow V_\mu+\der_\mu\t^1$. (The reason for the
superscript 1 will become clear shortly.)
As will be described, in the interacting theory, the tensor field
couples to certain Chern-Simons forms.  Closure of the algebra
then requires the vector transformation to act as well on the tensor field.
The precise form of this transformation will be discussed below.
It is necessary to first define the complete system of multiplets
which we wish to describe, and to define some notational
conventions.

We consider the vector-tensor multiplet in a background of
$n$ vector multiplets. One of these provides the gauge field
for the central charge, which we denote $W_\mu^0$. This must be an abelian
gauge field. The remaining $n-1$ vector multiplets supply additional
background gauge fields $W_\mu^a$, which need not be abelian.
The index $a$ is taken to run from $2$ to $n$.
We reserve the index $1$ for the vector field $V_\mu$ of the
vector-tensor multiplet.
(The reason for this choice is based on the dual description
of our theory, where the vector-tensor multiplet is replaced with
a vector multiplet, so that the dual theory involves $n+1$ vector
multiplets.) Also, since $W_\mu^0$ is the gauge field for the central charge,
the associated transformation parameter $\t^0$
is identified with the central charge parameter $z$ introduced above,
ie: $z\equiv\t^0$.  The vector gauge transformations act as follows
on the background gauge fields,
\bq \d W_\mu^0=\der_\mu z \,,\qquad
     \d W_\mu^a=\der_\mu\t^a+f^a_{\,bc}\t^bW_\mu^c\,.
\eq
As mentioned above, in the interacting theory,
the tensor field $B_{\mu\nu}$ necessarily couples to Chern-Simons forms.
This coupling is evidenced by the transformation behavior of the tensor.
To illustrate this, if we ignore the central charge (other than
its contribution to $W^0_\m$), then the
vector field of the vector-tensor multiplet would transform as
\bq \d V_\mu=\der_\mu\t^1
\label{simv}\eq
and the tensor field would transform as
\bq \d B_{\mu\nu}=2\der_{[\mu}\Lambda_{\nu]}
     +\eta_{IJ}\,\theta^I\der_{[\mu}W_{\nu]}^J\,,
\label{simb}\eq
where $\t^I$ and $\Lambda_\mu$ are the parameters of the transformations
gauged by $W_\mu^I$ and $B_{\mu\nu}$ respectively, and the index $I$
is summed from $0$ to $n$. As mentioned above, in this context
$W_\mu^1$ is identified with $V_\mu$.
Closure of the combined vector and tensor gauge
transformations requires that $\eta_{IJ}$ must be a constant tensor
invariant under the gauge group.
Furthermore, there is an ambiguity in the structure of
$\eta_{IJ}$, which derives from the possibility of performing
field redefinitions.  For example, without loss of generality,
$\eta_{IJ}$ can be chosen
symmetric by absorbing a term $\eta_{[IJ]}W_\mu^IW_\nu^J$ into
the definition of the tensor field $B_{\mu\nu}$.
This point illustrates a feature that plays an
important role in our results, namely that the presence of the
vector multiplets allows for background-dependent field redefinitions.
There is a similar issue for the other fields of
the vector-tensor multiplet, which can also be redefined by
terms depending on the background fields. We return to this issue
later. Using such redefinitions, we remove, without loss of generality,
all components of $\eta_{IJ}$ except for
$\eta_{11}, \eta_{1a}$ and $\eta_{ab}$,
and also render $\eta_{ab}$ symmetric.
Note that since $\eta_{1a}$ is invariant under the gauge
group, it follows that $\eta_{1a}W_\mu^a$ is an abelian
gauge field.

The situation is actually more complicated, since $V_\mu$ and $B_{\mu\nu}$
are also subject to the central charge transformation.
As described above, under this transformation these fields transform
into complicated dependent expressions, denoted $V_\mu^{({\rm z})}$ and
$B_{\mu\nu}^{({\rm z})}$ respectively, which involve other fields of
the theory.
Accordingly, we deform the transformation rule (\ref{simv}) to
\bq \d V_\mu = \der_\mu \t^1+z V_\mu^{({\rm z})} \,,
\eq
and, at the same time, (\ref{simb}) to
\bq \d B_{\mu\nu} = 2\der_{[\mu}\Lambda_{\nu]}
    +\eta_{11}\,\theta^1\der_{[\mu}V_{\nu]}
    +\eta_{1a}\,\theta^1\der_{[\mu}W_{\nu]}^a
    +\eta_{ab}\,\theta^a\der_{[\mu}W_{\nu]}^b
    +z B_{\mu\nu}^{({\rm z})} \,.
\eq
All $\t^0$-dependent terms, including any such Chern-Simons
contributions, are now included in $V_\m^{({\rm z})}$
and $B_{\mu\nu}^{({\rm z})}$, which are determined by closure of the full algebra, including supersymmetry.
Note that the central charge acts trivially on the components of the
vector multiplets, but not
on the fields of the vector-tensor multiplet.
The deformed transformation rules must still lead to a
closed gauge algebra, which requires $B_{\mu\nu}^{({\rm z})}$ to take the form
\bq B_{\mu\nu}^{({\rm z})} = -\eta_{11}V_{[\mu}V_{\nu]}^{({\rm z})}
    +\hat{B}_{\mu\nu}^{({\rm z})} \,,
\label{bzform}\eq
where $\hat{B}_{\mu\nu}^{({\rm z})}$ and $V_\mu^{({\rm z})}$ transform covariantly under the central charge, but
are invariant under all other gauge symmetries.  The resulting gauge algebra
now consists of the standard gauge algebra for the vector fields
augmented by a tensor gauge transformation.

To complete the discussion of the geometrical features of the
deformed gauge algebra, we list the fully covariant field strengths
for the vector and tensor gauge fields.
For the gauge fields of the background vector multiplets, we
retain the standard expressions,
\bq \F^0_{\mu\nu}=2 \der_{[\mu} W_{\nu]}^0 \,,\qquad
    \F^a_{\mu\nu}=2 \der_{[\mu} W_{\nu]}^a-f^a_{\,bc} W^b_\m
    W_\n^c\,,
\label{backfs}\eq
where we have included possible nonabelian corrections in
$\F_{\mu\nu}^a$.
As mentioned above, the gauge field associated with the central charge
must be abelian. For the gauge fields of the vector-tensor multiplet
the field strengths are
\brr F_{\mu\nu} &=& 2\der_{[\mu}V_{\nu]}
     -2W^0_{[\mu}V_{\nu]}^{({\rm z})}\,, \nonumber\\
     H^\mu &=& \ft12 i\ve^{\mu\nu\rho\s} \Big[\der_\nu B_{\rho \s}
     - \eta_{11} V_\nu \,\der_\rho V_\s
     - \eta_{1a} V_\nu \,\der_\rho W_\s^a \nonumber\\
     & & \hspace{1.05in}
     - \eta_{ab} W_\nu^a  \, (\der_\rho W_\s^b -
     \ft13 f^b_{\,cd}W_\rho^c
     W_\s^d) - W_\nu^0 \hat{B}_{\rho\s}^{({\rm z})}\Big] \,,
\label{fs}\err
which are covariant under the combined gauge transformations,
including those generated by the central charge.
The Bianchi identities corresponding to the field strengths
(\ref{backfs}) and (\ref{fs}) are straightforward to determine,
and are given by the following expressions,
\brr && \der_\mu\tilde{\F}^{0\mu\nu} = 0 \,,\qquad   D_\mu
     \tilde{\F}^{a\mu\nu} = 0  \,,\qquad  D_\mu\tilde{F}^{\mu\nu} =
     -V_\mu^{({\rm z})}\tilde{\F}^{0\mu\nu} \,,  \\
     &&  D_\mu H^\mu =- \ft18 i\ve^{\mu\nu\rho\s} \Big[\eta_{11}
     F_{\mu\nu} F_{\rho\s}
     +\eta_{1a} F_{\mu\nu} \F^a_{ \rho\s}
     +\eta_{ab} \F_{\mu\nu}^a \F^b_{\rho\s}
     +2\hat B_{\mu\nu}^{({\rm z})} \F^0_{\rho\s}  \Big]\,,
\nonumber
\label{bianchis}\err
where the derivative on the left-hand side is covariant with
respect to central-charge and nonabelian gauge transformations.

Observe that we have not yet specified the form of $\hat
B_{\mu\nu}^{({\rm z})}$, which is determined by supersymmetry
and will be discussed in the next section. As it turns out $\hat
B_{\mu\nu}^{({\rm z})}$ is itself proportional to the field
strengths (at least, as far as the purely bosonic terms are
concerned), but with field-dependent coefficients. Those
contributions thus characterize additional Chern-Simons terms
involving $W_\m^0$ which also involve the scalar fields.

We also draw attention to the fact that the Bianchi identity for
$H_\m$ is not linear in the vector-tensor fields. On the
right-hand side there are nonlinear terms that are either of
second-order (the term proportional to $\eta_{11}$) or of
zeroth-order (the term proportional to
$\eta_{ab}$) in the vector-tensor fields. Furthermore the
quantity $\hat B_{\mu\nu}^{({\rm z})}$ does not depend
homogeneously on the vector-tensor fields either. Hence,
generically the vector-tensor multiplet is realized in a
nonlinear fashion. One may attempt, by restricting the parameters
$\eta_{IJ}$ in a certain way, to find relatively simple
representations. However, supersymmetry severely restricts the
choices that one can make. In fact, as we intend to prove in the
next section, there are just two inequivalent vector-tensor
multiplets, associated with certain parameter choices.

\setcounter{equation}{0}
\section{The vector-tensor transformation rules}
Transformation rules can be determined by imposing the
supersymmetry algebra iteratively on multiplet component fields.
The supersymmetry transformation rules for vector multiplets are
fully known. Therefore, the algebra represented
by the vector-tensor multiplet in the presence of a vector
multiplet background is fixed up to gauge transformations
which pertain exclusively to the vector-tensor multiplet.
The most relevant commutator in this algebra
involves two supersymmetry transformations, which reads
\bq [\d_Q(\e_1),\d_Q(\e_2)] =
    \d^{({\rm cov})}(2\bar{\e}_2^i\g^\mu\e_{1i}+{\rm h.c.})
    +\d_{{\rm z}}(4\ve^{ij}\bar{\e}_{2i}\e_{1j}X^0+{\rm h.c.})
    +\d_{\rm g}(\t^1,\t^a,\Lambda_\mu) \,,
\label{algebra}
\eq
closing into a covariant translation $\d^{({\rm cov})}$,
a central charge transformation $\d_{{\rm z}}$, and vector and a tensor
gauge transformations (collectively denoted by $\d_{\rm g}$), each
with field-dependent parameters.
The field $X^0$ is the complex scalar of the vector multiplet
associated with the central charge\footnote{%
  Henceforth we will suppress the superscript on $X^0$ and define
  $X\equiv X^0$ to simplify the formulae.}. %
The gauge transformations in $\d_{\rm g}$ are found
by imposing (\ref{algebra}) on the vector-tensor multiplet.
For the final result one can verify that a supersymmetry transformation
and a gauge or central-charge transformation close into a gauge
or central-charge transformation. Gauge and central-charge
transformations form a subalgebra, which can be evaluated on the
basis of the results of the previous section.

We follow a procedure in which the vector multiplets play a
double role. As emphasized above, one of the vector multiplets
is required to realize the central charge in a local fashion.
In addition we require all couplings to be invariant under constant
scale and chiral $U(1)$ transformations.
These transformations act on the vector multiplet
components in a manner dictated by their behavior
as superconformal multiplets.
This leads to the scale and chiral weights
for the vector multiplet components
shown in table 1.  By insisting on invariance under scale and
chiral transformations we facilitate the coupling to supergravity
in the context of the superconformal multiplet calculus.
This topic will be discussed in a separate publication \cite{vtsg}.
The requirement of scale and chiral invariance may seem overly
restrictive.  However, this is not so,
since freezing some of the vector multiplets
to a constant leaves supersymmetry unaffected while at the same time
causing a breakdown of the scale and chiral symmetry.
\begin{figure}
\begin{center}
\begin{tabular}{|c||cccc||ccccc||c|}
\hline
&
\multicolumn{4}{c||}{vector multiplet} &
\multicolumn{5}{c||}{vector-tensor multiplet} &
\multicolumn{1}{c|}{parameter}  \\
\hline
\hline
field             &
   $X^I$          &
   $\O_i^I$       &
   $W_\mu^I$      &
   $Y_{ij}^I$     &
   $\phi$         &
   $V_\mu$        &
   $B_{\mu\nu}$   &
   $\l_i$         &
   $\phi^{({\rm z})}$   &
   $\e^i$           \\[.5mm]
\hline
\hline
$w$         & $1$  & $\ft32$  & $0$ & $2$     &
            $0$  & $0$      & $0$ & $\ft12$ & $0$ &
            $-\ft12$        \\[.5mm]
\hline
$c$         & $-1$ & $-\ft12$ & $0$ & $0$     &
            $0$  & $0$      & $0$ & $\ft12$ & $0$ &
            $-\ft12$       \\[.5mm]
\hline
$\gamma_5$&      & $+$      &     &         &
                 &          &     & $+$     &     &
            $+$        \\[.5mm]
\hline
\end{tabular}\\[.13in]
\parbox{5in}{Table 1: Scaling and chiral weights ($w$ and $c$, respectively)
             and fermion chirality ($\gamma_5$)
             of the vector and vector-tensor component fields
             and the supersymmetry parameter.}
\end{center}
\end{figure}

For the vector-tensor multiplet, there remains some flexibility
in the assignment of the scaling and chiral weights.
This is because the scalar fields of the vector multiplets
may serve as compensator fields. That is, we may arbitrarily
adjust the weights for each of the vector-tensor components by
suitably absorbing functions of $X^I$.  In this way we choose the
weights for the vector-tensor components to be as shown in
table 1.  The bosonic vector-tensor fields must all have chiral
weight $c=0$ since they
are all real.  In the context of supergravity, the scale and chiral transformations become local.  To eventually permit a consistent coupling
to supergravity, we must avoid a conflict between scale transformations and
vector and tensor gauge transformations. For this reason we adjust
$V_\mu$ and $B_{\mu\nu}$ to be also neutral under scale transformations.
The scale weights for $\phi$ and $\l_i$, and the scale and chiral weight
for $\l_i$ have been chosen somewhat arbitrarily.
There remains a freedom to absorb additional combinations
of the background fields into the definition of $\phi$ and $\l_i$.
Furthermore, the fields $V_\m$ and $B_{\mu\nu}$ can be redefined by
appropriate additive terms.
The freedom to redefine the vector-tensor components
by field redefinitions has already been mentioned in the previous section.
It is important to separate relevant terms in the
transformation rules from those that can be absorbed into such
field redefinitions. In deriving our results this aspect
has received proper attention.  We will shortly demonstrate some
of its consequences.

With diligence, solving the above
prescription is straightforward. In the following we suppress
nonabelian terms for the sake of clarity; they are not important
for the rest of this paper. We are not aware of arguments that would
prevent us from switching on the nonabelian interactions.
Also for the sake of clarity, we introduce a couple of abreviations,
for factors which occur frequently in what follows,
\bq g=i\eta_{1a}\frac{X^a}{X} \,,\qquad
    b=-\ft14 i\eta_{ab}\frac{X^aX^b}{X^2} \,.
\label{bgdef}\eq
We first list the transformation rules for arbitrary Chern-Simons
terms. Subsequently we will show that, in fact, they incorporate
two distinct versions of the vector-tensor multiplet.
The transformation rules for the vector-tensor multiplet
are given as follows,
\brr \d\phi &=& \bar{\e}^i\l_i
     +\bar{\e}_i\l^i \,,\nonumber\\
     \d V_\mu &=& i\ve^{ij}\bar{\e}_i\g_\mu
     \bpl 2X\l_j+\phi\O_j^0\bpr
     -iW_\mu^0\,\bar{\e}^i\l_i+ {\rm h.c.} \,, \nonumber\\
     \d B_{\mu\nu} &=& -2|X|^2\bar{\e}^i\s_{\mu\nu}\bpl
     4\eta_{11}\phi-g-\bar{g}\bpr\l_i \nonumber\\
     & & -2\bar{\e}^i\s_{\mu\nu}\bpl
     2\eta_{11}\phi^2\bar{X}\O_i^0
     +\phi\bar{X}^2\der_{\bar{I}}\bar{g}\,\O_i^I
     -4i\bar{X}{\rm Re}[\der_I(Xb)]\O_i^I \bpr\nonumber\\
     & & +\ve^{ij}\bar{\e}_i\g_{[\mu}V_{\nu]}\bpl
     2i\eta_{11}X\l_j
     +i\eta_{11}\phi\O_j^0
     +\eta_{1a}\O_j^a\bpr \nonumber\\
     & & +\ve^{ij}\bar{\e}_i\g_{[\mu}W_{\nu]}^0\bpl
     2X(2\eta_{11}\phi-g)\l_j
     +\eta_{11}\phi^2\O_j^0
     -i\eta_{1a}\phi\O_j^a
     -4i\der_I(Xb)\O_j^I\bpr \nonumber\\
     & & -i\eta_{11}W_{[\mu}^0V_{\nu]}\bar{\e}^i\l_i
     +\eta_{ab}\ve^{ij}\bar{\e}_i\g_{[\mu}W_{\nu]}^a\O_j^b+
     {\rm h.c.}\,,\nonumber\\
     \d\l_i &=&
     \bpl\Dslash\phi-i\Vslash^{({\rm z})}\bpr\e_i
     -\frac{i}{2X}\ve_{ij}\s\cdot\bpl F-i\phi\F^0\bpr\e^j
     +2\ve_{ij}\bar{X}\phi^{({\rm z})}\e^j \nonumber\\
     & & -\frac{1}{X}(\bar{\e}^j\l_j)\O_i^0
     -\frac{1}{X}(\bar{\e}^j\O_j^0)\l_i \\
     & & -\frac{1}{4\eta_{11}\phi-g-\bar{g}}\frac{1}{X}\bl
     \bpl 2\eta_{11}\phi^2Y_{ij}^0
     +\phi\bar{X}\der_{\bar{I}}\bar{g}\,Y_{ij}^I
     -4i{\rm Re}\,\der_I(Xb)Y_{ij}^I\bpr\e^j \nonumber\\
     & & \hspace{1.4in}
     -2\eta_{11}\e^j\bpl
     X\bar{\l}_i\l_j-\bar{X}\ve_{ik}\ve_{jl}\bar{\l}^k\l^l\bpr \nonumber\\
     & & \hspace{1.4in}
     +X\e^k(X\der_I g\,\bar{\O}_{(i}^I\l_{k)}
     -\bar{X}\ve_{il}\ve_{km}\der_{\bar{I}}\bar{g}\,
     \bar{\O}^{I(l}\l^{m)}) \nonumber\\
     & & \hspace{1.4in}
     +i\e^j(\der_I\der_J(Xb)\,\bar{\O}_i^I\O_j^J
     +\ve_{ik}\ve_{jl}\,\der_{\bar{I}}\der_{\bar{J}}(\bar{X}\bar{b})
     \bar{\O}^{Ik}\O^{Jl})\br\,.  \nonumber
\label{momrules}\err
Imposing closure of the supersymmetry algebra also
leads to the constraints on the higher elements of the
central charge hierarchy, as described in the previous secton.
This reflects the inability of the basic representation to
constitute an off-shell multiplet without central charge.
The following constraints are obtained in this manner,
\brr V^{({\rm z})}_\mu &=&
     \frac{-1}{4\eta_{11}\phi-g-\bar{g}}
     \frac{1}{|X|^2}\bl
     H_\mu-[i(2\eta_{11}\d_I\,^0\phi^2+\phi X\der_I g-4{\rm Im}\der_I(Xb))
     X\der_\mu\bar{X}^I+{\rm h.c.}]\br \nonumber\\
     & & + {\rm fermion\ terms}\,,\nonumber\\[.1in]
     \hat{B}^{({\rm z})}_{\mu\nu}&=&
     \ft14 i(g-\bar{g})F_{\mu\nu}
     +\ft14 i(4\eta_{11}\phi-g-\bar{g})\tilde{F}_{\mu\nu}
     -\ft14\phi(2\eta_{11}\phi-g-\bar{g})\F_{\mu\nu} \nonumber\\
     & & +\ft12 i\phi{\rm Im}(X\der_I g)\tilde{\F}_{\mu\nu}^I
     +2{\rm Im}[\der_I(Xb)](\F_{\mu\nu}^I-\tilde{\F}_{\mu\nu}^I) 
     + {\rm fermion\ terms} \,,
\label{vzbz}\err
as well as similar relations for $\l^{({\rm z})}$ and $\phi^{({\rm zz})}$,
which are of less direct relevance. The specific constraints
shown in (\ref{vzbz}) are crucial for determining
supersymmetric couplings, as we discuss in subsequent sections.
By acting on the above
constraints with central-charge transformations, one recovers
an infinite hierarchy of
constraints.  These relate the components of the
higher multiplets
$(\phi^{({\rm z})}, V_\mu^{({\rm z})}, B_{\mu\nu}^{({\rm z})}, \l_i^{({\rm z})},
\phi^{({\rm zz})})$, etcetera, in such a way as to
retain precicely $8+8$ independent degrees of freedom.
The transformation rules (\ref{momrules}) are completely general,
in the sense that no modifications are possible which cannot be removed by
field redefinitions.  These, then, constitute a
unique representation of the vector-tensor multiplet. As
exhibited in \cite{vt1} the transformation rules of the higher
multiplets take almost the same form. Their characteristic
feature is that the transformations involve objects both at the
next and at the preceding level. The transformations given above
involve only the next level as there is no lower level. The
consistency of this is ensured by the gauge transformations of
the fields $V_\m$ and $B_{\m\n}$.

At this point we come to a crucial feature of our results. We will
now demonstrate that, depending on whether $\eta_{11}$ vanishes or
not, i.e. depending on whether or not we include a $V\wedge{\rm d}V$
Chern-Simons form in the tensor couplings, we describe
two {\it inequivalent} representations of the vector-tensor
multiplet.
\vspace{.2in}

\noindent{\it The nonlinear vector-tensor multiplet}:\\
If $\eta_{11}\not=0$, signifying the presence of a
$V\wedge{\rm d}V$ Chern-Simons form in the tensor couplings,
then we can redefine fields in such a way that
$\eta_{1a}$ disappears completely from the formulation described so far,
signifying the absence of $V\wedge{\rm d}W^a$
Chern-Simons forms in the tensor couplings.
Specifically, if we begin with the transformation rules (\ref{momrules}),
and if $\eta_{11}\ne 0$, then we may perform the following redefinition,
\brr \phi &\longrightarrow& \phi
     -\ft14 i\frac{\eta_{1a}}{\eta_{11}}
     \Big(\frac{X^a}{X}-\frac{\bar{X}^a}{\bar{X}}\Big)\,, \nonumber\\
     V_\mu &\longrightarrow& V_\mu
     -\ft14\frac{\eta_{1a}}{\eta_{11}}
     \Big(\frac{X^a}{X}+\frac{\bar{X}^a}{\bar{X}}\Big)W_\mu^0
     +\ft12\frac{\eta_{1a}}{\eta_{11}}W_\mu^a\,, \nonumber\\
     B_{\mu\nu} &\longrightarrow& B_{\mu\nu}
    +\ft14\eta_{1a}\Big(\frac{X^a}{X}+\frac{\bar{X}^a}{\bar{X}}\Big)
     V_{[\mu}W_{\nu]}^0
     +\ft12\eta_{1a}V_{[\mu}W_{\nu]}^a \nonumber\\
     & & \hspace{.34in}
     -\ft1{16}\frac{\eta_{1a}\eta_{1b}}{\eta_{11}}
     (\frac{X^b}{X}-\frac{\bar{X}^b}{\bar{X}})
     W_{[\mu}^0W_{\nu]}^a \,.
\label{redef}\err
In terms of the shifted fields, we then obtain precisely the
rules (\ref{momrules}), but without the $\eta_{1a}$ terms.  This
version of the vector-tensor multiplet is a straightforward
extension of the result presented in \cite{vt1}, but with the
background extended to several vector multiplets. A
characteristic feature of this version is that the transformation
rules are nonlinear in the vector-tensor components, as a result
of the Chern-Simons coupling between $V_\m$ and $B_{\m\n}$.
Observe that this version contains at least two abelian vector
gauge fields, $W^0_\m$ and $V_\m$.
\vspace{.2in}

\noindent{\it The linear vector-tensor multiplet}:\\
If $\eta_{11}=0$, signifying the absence of a
$V\wedge{\rm d}V$ Chern-Simons coupling,
thereby avoiding nonlinearities (in terms of vector-tensor fields)
in the Bianchi identities (\ref{bianchis}), we arrive at a
distinct formulation.
This case has at least three abelian vector fields, $W^0_\m$,
$\eta_{1a}W_\m^a$ and $V_\m$. As one can check from
(\ref{momrules}), all the transformation rules now become
at most linear in the fields of the vector-tensor multiplet.
\setcounter{equation}{0}
\section{Linear Multiplet Construction}
It is possible to form products of vector tensor-multiplets,
using the background vector multiplets judiciously, so as to form
$N=2$ linear multiplets.
This enables the construction
of supersymmetric actions using known results in multiplet
calculus.
In this section we describe the construction
of such linear multiplets. In the following section we present
the associated supersymmetric actions, and give the dual descriptions
in terms of vector multiplets alone.

A linear multiplet has $8+8$ independent off-shell degrees of freedom.
It comprises a triplet of real scalars, $L_{ij}$,
which satisfy $L_{ij}=\ve_{ik}\ve_{jl}L^{kl}$,
a fermion doublet $\varphi_i$, a complex scalar $G$, and a real vector
field $E_\mu$.  It can support a nontrivial central charge,
which would then generate an infinite hierarchy of multiplets,
supplemented by constraints, in a manner completely analogous
to the vector-tensor multiplet, discussed above.
The component fields transform under supersymmetry as
follows \cite{DWLVP},
\brr \d L_{ij} &=& 2\bar{\e}_{(i}\vphi_{j)}
     +2\ve_{ik}\ve_{jl}\bar{\e}^{(k}\vphi^{l)}\,, \nonumber\\
     \d\vphi^i &=& \Dslash L^{ij}\e_j+\Eslash\ve^{ij}\e_j-G\e^i
     +2\bar{X}L^{({\rm z})ij}\ve_{jk}\e^k\,, \nonumber\\
     \d G &=& -2\bar{\e}_i\Dslash\vphi^i
     +2\bar{X}\big(\ve^{ij}\bar{\e}_i\vphi_j^{({\rm z})}-h.c.\big)
     -2\bar{\e}_i\O^j L^{({\rm z})ik}\ve_{jk}\,, \nonumber\\
     \d E_\mu &=& 2\ve_{ij}\bar{\e}^i\s_{\mu\nu}D^\nu\vphi^j
     +2\bar{X}\bar{\e}^i\g_\mu\vphi_i^{({\rm z})}
     +\bar{\e}^i\g_\mu\O^j L_{ij}^{({\rm z})}+{\rm h.c.} \,,
\label{linear}\err
where $L_{ij}^{({\rm z})}$ and $\varphi_i^{({\rm z})}$ are
the image of $L_{ij}$ and $\varphi_i$ under the central charge.
We stress that the fields $X$ and $\O_i$ which appear
in (\ref{linear}) are the scalar and fermion doublet,
respectively, of the vector multiplet which contains
the gauge field for the central charge transformation,
referred to as $X^0$ and $\O_i^0$ above.
As usual, we suppress the superscript zeros for the sake
of clarity.
As mentioned, objects higher in the central-charge hierarchy, 
like $L_{ij}^{({\rm z})}$ and
$\varphi_i^{({\rm z})}$ are dependent.  Their exact form is
not particularly relevant because we construct a linear
multiplet as wholly dependent on vector-tensor and vector multiplet
components.
\begin{figure}
\begin{center}
\begin{tabular}{|c||cccc|}
\hline
function &${\cal A}$&
${\cal B}_I$&
${\cal C}_{IJ}$&
${\cal D}_I$\\
\hline
\hline
w & $0$ & $-1$ & $-1$ & $0$ \\
\hline
c & $0$ & $1$  & $1$  & $0$ \\
\hline
\end{tabular}\\[.13in]
\parbox{2.5in}{Table 2: Scaling and chiral weights ($w$ and $c$, respectively)
             of the functions ${\cal A}, {\cal B}_I, {\cal C}_{IJ}$
and ${\cal D}_I$.}
\end{center}
\end{figure}

We determine the linear multiplet by requiring the
lowest component $L_{ij}$ to have weights $w=2$ and $c=0$ and
to have the suitable transformation property.
We wish to construct $L_{ij}$ as a function of the
vector-tensor fields and the background vector multiplet fields.
To discover the form of this construction, we notice that the
$L_{ij}$ must transform nontrivially under chiral $SU(2)$
transformations.  The only vector-tensor component
which transforms under $SU(2)$ is the fermion $\l_i$.
For the vector multiplets, only the fermions $\O_i^I$ and
the auxiliary fields $Y_{ij}^I$ transform nontrivially under $SU(2)$.
Therefore, the most general possible linear multiplet must include
an $L_{ij}$ of the following form
\brr L_{ij} &=& X{\cal A}\,\bar{\l}_i\l_j
     +\bar{X}\bar{{\cal A}}\,\ve_{ik}\ve_{jl}\bar{\l}^k\l^l \nonumber\\
     & & +X{\cal B}_I\,\bar{\l}_{(i}\O_{j)}^I
     +\bar{X}\bar{{\cal B}}_{\bar{I}}\,
     \ve_{ik}\ve_{jl}\bar{\l}^{(k}\O^{Jl)} \nonumber\\
     & & +{\cal C}_{IJ}\,\bar{\O}_i^I\O^J_j
     +\bar{{\cal C}}_{\bar{I}\bar{J}}\,
     \ve_{ik}\ve_{jl}\bar{\O}^{Ik}\O^{Jl} \nonumber\\
     & & +{\cal D}_IY_{ij}^I \,,
\label{ansatz}\err
where ${\cal A}, {\cal B}_I, {\cal C}_{IJ}$
and ${\cal D}_I$ are functions of $\phi,\, X^I$ and
$\bar{X}^I$, with scaling and chiral weights as shown in table 2.
Equation (\ref{ansatz}) is the natural ansatz to begin a
systematic analysis.

Requiring that $L_{ij}$ transform into a spinor doublet as indicated in
(\ref{linear}) puts stringent requirements on each of the functions
${\cal A}(\phi,X^I,\bar{X}^I)$,
${\cal B}_I(\phi,X^I,\bar{X}^I)$,
${\cal C}_{IJ}(\phi,X^I,\bar{X}^I)$
and ${\cal D}_I(\phi,X^I,\bar{X}^I)$.
These are encapsulated by a system of coupled first-order, linear differential
equations, which are determined as follows.
Upon varying (\ref{ansatz}) with respect to supersymmetry,
one finds that the resulting 3-fermi terms and terms involving $Y_{ij}^I$
take the required form if and only if the following conditions
are satisfied,
\brr {\cal E}\der_\phi{\cal A} &=&
     -4\eta_{11}\bar{{\cal A}}\,, \hspace{1.4in}
     {\cal E}\der_{\bar{I}}{\cal B}_J =
     \bar{{\cal B}}_{\bar{I}}\der_J g\,, \nonumber\\
     {\cal E}\der_I{\cal A} &=&
     ({\cal A}+\bar{{\cal A}})\der_I g
     +2\eta_{11}{\cal B}_I\,, \hspace{.4in}
     {\cal E}\der_\phi{\cal C}_{IJ} =
     2 i\bar{{\cal A}}\der_I\der_J(Xb)\,,\nonumber\\
     {\cal E}\der_{\bar{I}}{\cal A} &=&
     -2\eta_{11}\bar{{\cal B}}_{\bar{I}}, \hspace{1.3in}
     {\cal E}\der_{\bar{K}}{\cal C}_{IJ} =
     i\bar{{\cal B}}_{\bar{K}}\der_I\der_J(Xb)\,, \nonumber\\
     {\cal E}\der_\phi{\cal B}_I &=&
     2\bar{{\cal A}}\der_I g\,, \hspace{1.6in}
     \der_\phi {\cal D}_{I} =
     -X{\cal B}_I
     -2{\cal A}P_I\,, \nonumber\\
     \der_{(I}(X^2 {\cal E}{\cal B}_{J)}) &=&
     4 i({\cal A}+\bar{{\cal A}})X\der_I\der_J(Xb)\,, \hspace{.5in}
     \der_I{\cal D}_J =
     -2{\cal C}_{IJ}
     -{\cal B}_IP_J\,,
\label{constraints}\err
where $g$ and $b$ were defined in (\ref{bgdef}) and
\brr {\cal E} &=& -4\eta_{11}\phi+g+\bar{g} \,, \nonumber\\
     P_I &=& -\ft12\phi\d_I\,^0
     -i{\cal E}^{-1}{\rm Im}\Big(
     \phi X\der_I g
     +4i\der_I(Xb)\Big) \,.
\label{epdef}\err
Furthermore, the reality condition on $L_{ij}$ requires that
${\cal D}_I$ be real.  It is satisfying that the
system of equations (\ref{constraints}) turns out
to be integrable, despite its complexity. After some work,
one can prove that the general solution decomposes as a
linear combination of three distinct solutions, each with
an independent physical interpretation. The most interesting of these
is given as follows,
\brr [{\cal A}{}]_1 &=&
     \eta_{11}(\phi+i\zeta)
     -\ft12 g \,, \nonumber\\
     {}[{\cal B}_I{}{}]_1 &=&
     -\ft12(\phi+i\zeta)\der_I g
     -2i\der_I b \,,\nonumber\\
     {}[{\cal C}_{IJ}]_1 &=&
     -\ft12 i(\phi+i\zeta)\der_I\der_J(Xb) \,, \nonumber\\
     {}[{\cal D}_I{}]_1 &=& {\rm Re}\bl
     [\ft13\eta_{11}(\phi+i\zeta)^3
     -\ft12 i\zeta(\phi+i\zeta)g]\d_I\,^0
     \nonumber\\
     & & \hspace{.3in}
     +\ft12(\phi+i\zeta)X\der_I(g\phi+4ib)\br \,,
\label{first}\err
where
\bq \zeta=\frac{{\rm Im}(\phi g+4ib)}
    {2\eta_{11}\phi-{\rm Re}\,g} \,.
\label{zetadef}\eq
In terms of the action, which is discussed in the next section,
this solution provides the couplings which involve the
vector-tensor fields.  The remaining two solutions,
which we discuss presently, give rise either to a total divergence
or to interactions which involve only the background fields.
The latter of these
correspond to previously known results.  For this reason, the
remaining solutions are secondary to that
presented in (\ref{first}). We discuss them to be complete,
but also to demonstrate the pervasive nature
of the techniques which we employ.

In addition to (\ref{first}), a second solution to
(\ref{constraints}) is given as follows,
\brr {}[{\cal A}{}]_2 &=&
     i\eta_{11}\zeta'
     -i\a \,, \nonumber\\
     {}[{\cal B}_I{}]_2 &=&
     -\ft12 i\zeta'\der_I g
     -2i\der_I\g \,,
     \nonumber\\
     {}[{\cal C}_{IJ}{}]_2 &=&
     \ft12\zeta'\der_I\der_J(Xb)\,, \nonumber\\
     {}[{\cal D}_I{}]_2 &=& {\rm Re}\bl
     2i X \phi \der_I \g + \ft i2 \z' X \phi \der_I g - 2 \z' \der_I
     (Xb)\br\,,
\label{second}\err
where $\g=\ft14 i\a_a{X^a}/{X}$, which is a holomorphic,
homogeneous function of the background scalars $X^a$
and $X^0$,
\bq \zeta'=\frac{2\a\phi+4{\rm Re}\,\g}
    {2\eta_{11}\phi-{\rm Re}\,g} \,,
\label{zpdef}\eq
and where $\a$ and $\a_a$ are arbitrary real parameters.
Note that this solution could be concisely included with the first
solution by redefining $g\to g+2i\a$ and $b\to b+\g$.
In fact, this second solution indicates that
the functions $g$ and $b$ are actually
defined modulo the modifications indicated by these shifts.
In terms of the action, this ambiguity is analogous
to the shift of the theta angle in an ordinary Yang-Mills theory.
The third and final solution to (\ref{constraints})
is given by
\brr {}[{\cal A}{}]_3 &=& 0 \,,
     \nonumber\\
     {}[{\cal B}_I{}]_3 &=& 0 \,,
     \nonumber\\
     {}[{\cal C}_{IJ}{}]_3 &=&
     -\ft18 i\der_I\der_J(X^{-1}f) \,,
     \nonumber\\
     {}[{\cal D}_I{}]_3 &=&
     -\ft12{\rm Im}\der_I\bpl X^{-1}f \bpr \,,
\label{third}\err
where $f$ is an arbitrary holomorphic function of $X^0$ and $X^a$,
of degree two. In terms of the action, this solution
corresponds to interactions amongst the background vector
multiplets alone.  Since the possible vector multiplet
self-couplings have been fully classified, this solution
does not provide us with new information.  At the same time,
however, it is reassuring and satisfying that the
previously established solutions
have been found anew in the present context.  The function $f$
provides the well-known holomorphic prepotential for describing
the background self-interactions.

Now that we have determined the scalar triplet $L_{ij}$,
in terms of the specific functions
${\cal A}(\phi,X^I,\bar{X}^I)$,
${\cal B}_I(\phi,X^I,\bar{X}^I)$,
${\cal C}_{IJ}(\phi,X^I,\bar{X}^I)$, and
${\cal D}_I(\phi,X^I,\bar{X}^I)$
given above, it is straightforward to generate
the remaining components of the linear multiplet,
$\varphi_i,\, G$, and $E_\mu$ by varying (\ref{ansatz})
with respect to supersymmetry.
The precise functional form of these higher components,
in terms of the vector-tensor and the background fields, is
not so illuminating, so we will not present them here.
Given the complete linear multiplet, it is straightforward
to determine a supersymmetric action.
This is discussed in the following section.
\section{Vector-tensor Lagrangians and their dual versions}
As mentioned above, there are known results
which describe supersymmetric densities as multiplet products.
Particularly useful is a product between an $N=2$ linear
multiplet and an $N=2$ vector multiplet which yields such a density,
\bq {\cal L}=
    -W^\mu E_\mu
    -\ft12 Y^{ij}L_{ij}
    +\{(XG+\bar{\O}_i\vphi^i)+{\rm h.c.}\} \,,
\label{product}\eq
where $L_{ij}$, $\varphi_i$, $G$, and $E_\mu$
are the components of a linear multiplet,
and where $X$, $W_\mu$, $\O_i$ and $Y_{ij}$ are the components of the
vector multiplet used to gauge the central charge, which
appear explicitly in the linear multiplet transformation
rules (\ref{linear}).  Again, we suppress the superscript zero in the interest of
clarity.
Equation (\ref{product}) represents a supersymmetric
Lagrangian for a generic linear multiplet.
We choose the linear multiplet in (\ref{product}) to be the one
discussed in the previous section.
In this way we are able to construct an interacting
supersymmetric Lagrangian involving the vector-tensor component
fields, which also respects the central charge as a local symmetry.
This explains the utility of constructing the
linear multiplet in the previous section.

In this paper we restrict
attention to the bosonic terms in the action, since these are
of the most interest. The terms involving fermions are
straightforward to generate, however.
Looking at equation (\ref{ansatz}), we
see that $L_{ij}$ consists of terms quadratic in fermions,
with one additional $Y_{ij}^I$ term.
From this we deduce the following.  From the second term
of (\ref{product}) only the $Y_{ij}^I$ term in (\ref{ansatz})
contributes to the bosonic Lagrangian, since the balance
of $L_{ij}$ contributes only fermion bilinears.  Therefore,
the second term of (\ref{product}) supplies only a term
$-\ft12{\cal D}_IY^0_{ij}Y^{ijI}$ to the bosonic Lagrangian.
The bulk of the bosonic Lagrangian then comes from
the $G$ and $E_\mu$ components of the linear multiplet.
From (\ref{linear}) we see that $G$ and $E_\mu$ are generated
by varying $L_{ij}$ {\it twice} with respect to supersymmetry.
Therefore, to obtain the purely bosonic action we need only consider
the variation of the fermion fields in each of the fermion
bilinear terms of $L_{ij}$. Any additional contributions will
neccesarily involve fermions, since each of the two variations of
$L_{ij}$ needed to generate $G$ and $E_\mu$ must remove
one of the two fermions from the respective term of $L_{ij}$.
The contributions which follow from the $Y_{ij}$ term in
(\ref{ansatz}) needs to be treated differently; there, one has to
consider seperately the second variation of ${\cal D}_I$ and
also that of
$Y_{ij}$ in order to generate the purely bosonic contributions
to $G$ and $E_\mu$.  Finally, the $\O_i\varphi^i$ term in
(\ref{product}) obviously does not contribute to the
bosonic action at all, so we ignore it.

Given the complexity of the
transformation rule for $\l_i$ found in (\ref{momrules}),
it is clear that a fair amount of work is involved in
carrying out this process.  Nevertheless, it is straightforward to
vary (\ref{ansatz}) to generate $\varphi_i$, $G$, and $E_\mu$,
which can then be read of from (\ref{linear}), then to insert these
expressions into the action formula (\ref{product}), in precisely
the manner described above. Carrying out
this process, one finds the following form for the bosonic action,
\brr {\cal L} &=&
     |X|^2{\cal A}\,(\der_\mu \phi -iV^{({\rm z})}_\mu)^2
     +2|X|^2{\cal B}_I\,\der^\mu X^I
     (\der_\mu\phi -iV_\mu^{({\rm z})}) \nonumber\\
     & & +4\bar X{{\cal C}}_{IJ}\,
     \der^\mu X^I\der_\mu X^J
     -2{\cal D}_I\, X\Box\bar{X}^I \nonumber\\
     & & +{\cal A}\,(F^{-\, \mu\nu}-i\phi\F^{-\, 0\mu\nu})\bpl
     \ft14(F_{\mu\nu}^--i\phi\F^{-\, 0}_{\mu\nu})
     +iW^{0}_{\mu}(\der_\nu \phi -iV_\nu^{({\rm z})})\bpr \nonumber\\
     & & +iX{\cal B}_I\,\F^{-\, I \mu\nu}\bpl
     \ft12(F_{\mu\nu}^--i\phi\F^{-\, 0}_{\mu\nu})
     +iW^0_\mu (\der_\nu \phi - iV_\nu^{({\rm z})})\bpr \nonumber\\
     & & +i{\cal B}_I\,(F^{-\, \mu\nu} - i\phi \F^{-\, \mu\nu})W^0_\mu
     \der_\nu X^I\nonumber\\
     & & -{\cal C}_{IJ}\,\F^{I-\mu\nu}\bpl X \F^{J-}_{\mu\nu} + 4 W^0_\mu
     \der_\nu X^J\bpr
     +{\cal D}_IW^{0\mu}\der^\nu\F_{\mu\nu}^{-\, I} \nonumber\\
     & & +|X|^2 {\cal A}\,(W_\mu^2 + 4|X|^2)(\phi^{({\rm z})})^2 \nonumber\\
     & & -\ft14(X\der_{(I}{\cal D}_{J)}
     +P_{(I}\der_\phi{\cal D}_{J)})Y_{ij}^IY^{Jij}
     -\ft14{\cal D}_I\,Y_{ij}^0 Y^{Iij}\nonumber\\
     & & +{\rm h.c.}
\label{lagrangian}\err
This describes the bosonic coupling of a vector-tensor multipet to
$n$ vector multiplets. Note that each term involves a factor
of one of the functions ${\cal A}(\phi,\,X^I,\,\bar{X}^I)$,
${\cal B}_I(\phi,\,X^I,\,\bar{X}^I)$,
${\cal C}_{IJ}(\phi,\,X^I,\,\bar{X}^I)$
${\cal D}_I(\phi,\,X^I,\,\bar{X}^I)$
or $P_I(\phi,\,X^I,\,\bar{X}^I)$, which were given explicitly
in the previous section.  It would be desirable to utilize
the precise forms of these functions, as well as the
properties (\ref{constraints}) to cast this Lagrangian
in a more elegant fashion, but this is not essential for
the purposes of this paper.
A special case of (\ref{lagrangian})
was presented previously in \cite{vt1}, where the background
consisted of a single vector
multiplet. In that case $\eta_{1a}$ and $\eta_{ab}$
were necessarily zero, so that the functions $g$ and $b$
were vanishing.

As described in the previous section, the functions
${\cal A}(\phi,\,X^I,\,\bar{X}^I)$,
${\cal B}_I(\phi,\,X^I,\,\bar{X}^I)$,
${\cal C}_{IJ}(\phi,\,X^I,\,\bar{X}^I)$ and
${\cal D}_I(\phi,\,X^I,\,\bar{X}^I)$,
which define the Lagrangian are linear superpositions
of three distinct terms, one of which describes the
local couplings of the vector-tensor multiplet components,
another which is a total derivative,
and one which codifies the self-interactions of the background.
As a result of this, the Lagrangian (\ref{lagrangian}) can
be written as a sum of three analogous pieces: a vector-tensor
piece, a total-derivative piece, and a piece that exclusively depends
on the background.

It would be interesting to use our Lagrangian (\ref{lagrangian})
to address physical questions, such as the quantum mechanical
properties of $N=2$ supersymmetric theories involving tensor
fields. As discussed in the introduction, it is expected that the
restrictions on the couplings inferred by tensor gauge fields
presumably implies special renormalization theorems,
perhaps even finiteness. 
Having Lagrangians like (\ref{lagrangian}) (including the
fermionic additions which have been suppressed in this paper,
but which can be obtained straightforwardly using methods
presented above) is imperative for addressing such issues.

In addition to the quantum aspects just mentioned,
there exist more humble but nevertheless relevant
questions which may be addressed with relative
ease.  For instance,
a vector-tensor multiplet is classically equivalent to a vector
multiplet. The theory which we have presented, involving
one vector-tensor multiplet and $n$ vector multiplets
is classically equivalent to a theory involving $n+1$ vector
multiplets. Since these latter theories are well understood, it
is of interest to determine what subset of vector multiplet
theories are classically equivalent to vector-tensor theories.
Furthermore, heterotic low-energy string Lagrangians with $N=2$
supersymmetry are usually based on vector multiplets rather than
on vector-tensor multiplets, despite the fact that the dilaton
in a heterotic string theory resides in a vector-tensor
multiplet.  This is justified
by the classical equivalence just mentioned.

A significant restriction along these lines has to do with the
K\"ahler spaces on which the scalar fields of the theory may live.
In the case of $N=2$ vector multiplets these consist of so-called
``special K\"ahler" spaces, and the attendent geometry is
known as ``special geometry". For the case of effective
Lagrangians corresponding to heterotic $N=2$ supersymmetric
string compactifications, a well-known theorem \cite{FVP}
indicates that this space must involve, at least at the string
tree level,
an independent $SU(1,1)/U(1)$ coset factor to accomodate
the axion/dilaton complex.
It would be interesting to discover such a restriction
from supersymmetry in a field theoretic context,
such as having it follow from the requirement
of a dual relationship with a vector-tensor theory. Such questions
had lead us to suppose that the vector-tensor theories had
dual formulations which uniformly posessed such a factorization.
As we discuss below, this has proved not to be the case.
Irrespective of this, it is useful to cast our
Lagrangian in terms of its dual form,
since there the couplings are concisely encoded in a single
holomorphic function.

One goes about constructing the dual formulation,
in the usual manner, by introducing a Lagrange multiplier
field $a$, which, upon integration, would enforce the Bianchi identity
on the field strength $H_\mu$.
The relevant term to add to the Lagrangian is therefore
\brr {\cal L}(a) &=&
     a(D_\mu H^\mu
     +\ft14 i\eta_{11}F_{\mu\nu}\tilde{F}^{\mu\nu}
     +\ft14 i\eta_{1a}F_{\mu\nu}\tilde{\F}^{a\mu\nu}
     +\ft14 i\eta_{ab}\F_{\mu\nu}^a\tilde{\F}^{b\mu\nu}
     +\ft12 i\hat{B}_{\mu\nu}^{({\rm z})}\tilde{\F}^{0\mu\nu})\,.
     \nonumber\\
\label{aterm}\err
Including this term, we treat $H_\mu$ as unconstrained and integrate
it out of the action, thereby trading the single on-shell degree
of freedom represented by $B_{\mu\nu}$ for the real scalar $a$.
Doing this, and also eliminating the auxiliary fields,
$\phi^{({\rm z})}$ and $Y_{ij}^I$, we obtain a dual theory involving only
vector multiplets.  To perform these operations, it is instructive to
note that all occurances of $H_\mu$ in
(\ref{lagrangian}) and (\ref{aterm}) are most conveniently written
in terms of $V_\mu^{({\rm z})}$, which can be done using (\ref{vzbz}).
All such terms can be then be collected, and written as follows,
\bq {\cal L}(V_\mu^{({\rm z})})=\ft14 {\cal E}\bpl
    W^\mu W^\nu-(W_\mu^2+4|X|^2)\d^{\mu\nu}\bpr
    \bpl -\ft12 V_\mu^{({\rm z})}V_\nu^{({\rm z})}
   +V_\mu^{({\rm z})}\der_\nu(a-\zeta)\bpr   \,,
\label{lvz}\eq
where ${\cal E}$ was defined in (\ref{epdef}) and
$\zeta$ was defined in (\ref{zetadef}).
It is interesting how the terms involving $V_\mu^{({\rm z})}$ factorize
into the form given in (\ref{lvz}). The equation of motion for $H_\mu$
is conveniently written in terms of $V_\mu^{({\rm z})}$.
From (\ref{lvz}), this follows immediately, and is given by
the following simple expression,
\bq V_\mu^{({\rm z})}=\der_\mu(a-\zeta)\,.
\eq
The equations of motion for the auxiliary fields are $\phi^{({\rm z})}=0$, and
$Y_{ij}^I=0$. After substituting these solutions, we then manipulate the
result into the familiar form for the bosonic Lagrangian
involving vector multiplets,
\bq {\cal L} = \ft12 i\big(\der_\mu F_I\,\der^\mu\bar{X}^I
     -\der_\mu X^I\,\der^\mu\bar{F}_I\big)
     -\ft18 i\big(\bar{F}_{IJ}\,\F^{+I}_{\mu\nu}\F^{+\mu\nu J}
     -F_{IJ}\,\F^{-I}_{\mu\nu}\F^{-\mu\nu J}\big)\,,
 \label{aform}\eq
characterized by a holomorphic function $F(X^0,X^1,X^a)$.
In (\ref{aform}), a subscript $I$ denotes differentiation
with respect to $X^I$.
The natural bosons in the dual theory are found to be
\brr X^1 &=& X^0\Big((a-\zeta)+i\phi\Big)\,, \nonumber\\
     W_\mu^1 &=& V_\mu+(a-\zeta)W_\mu^0 \,.
\err
One can check that these transform as components
of a common vector multiplet.
For the general case, the dual theory obtained
in this manner is described by the following
holomorphic prepotential,
\brr F &=& -\frac{1}{X^0}\bpl
    \ft13\eta_{11}X^1X^1X^1
    +\ft12\eta_{1a}X^1X^1X^a
    +\eta_{ab}X^1X^aX^b \bpr \nonumber\\
    & & -\a X^1X^1
    +\a_a X^1X^a
    +f(X^0,X^a) \,.
\label{prepot}\err
The quadratic terms proportional to $\a$ and $\a_a$
give rise to total derivatives since
their coefficients are real.  The term involving
the function $f(X^0,X^a)$ represents the self-interactions
of the background vector multiplets.
The first three terms in (\ref{prepot})
encode the couplings of the erstwhile vector-tensor fields,
$\phi$ and $a$, and it is these which we are most interested in.
As mentioned above, it is natural to question whether the
K\"ahler space described by this prepotential function
conforms to the theorem of \cite{FVP}.  The fact that it does not
derives from the fact that the would-be complex dilaton
$X^1/X^0$ appears more than linearly in the prepotential.
Nor can the quadratic or cubic terms be removed by a clever
field redefinition.  As discussed earlier in this paper, the
best one can do it to remove {\it either} $\eta_{11}$ or
$\eta_{1a}$.  There exists an obstruction to removing both
of these.  We remark that these parameters are related to the
Chern-Simons couplings to the tensor field in the dual
formulation.  The obstruction to removing the unwanted terms
in the prepotential derive from the inability to formulate
an interacting vector-tensor theory without any such Chern-Simons
couplings.

It is important to realize that these results are
a concise description of two very different situations.  As described
in detail in section 3, depending on whether the parameter
$\eta_{11}$ is vanishing or not, indicating the absence or presence,
respectively, of a $V\wedge {\rm d}V$ Chern-Simons coupling to the tensor
field, the theory takes on very distinct characters.  It is instructive
then, to summarize our results independently for each of these two
cases.
\vspace{.3in}

\noindent{\it The nonlinear vector-tensor multiplet:}\\[.05in]
As described above, when the parameter $\eta_{11}$ does not vanish,
the tensor field involves a coupling to the Chern-Simons form
$V\wedge {\rm d}V$, which is quadratic in terms of vector-tensor fields.
Consequently, the corresponding transformation rules contain
significant nonlinearities.  As also discussed above, in this case
it is possible to remove the parameter $\eta_{1a}$, and therefore
the $V\wedge {\rm d} W^a$ Chern-Simons couplings, without loss of
generality, by the field redefinition given in (\ref{redef}).
Without loss of generality, we then define
$\eta_{11}=1$ and $\eta_{1a}=0$. In this case the functions
${\cal A}(\phi,X^I,\bar{X}^I)$,
${\cal B}_I(\phi,X^I,\bar{X}^I)$,
${\cal C}_{IJ}(\phi,X^I,\bar{X}^I)$, and
${\cal D}_I(\phi,X^I,\bar{X}^I)$
which define the linear multiplet and, more importantly,
the vector-tensor Lagrangian (\ref{lagrangian}) are given by
the following expressions
\brr {\cal A} &=& \phi+i\frac{b+\bar{b}}{\phi}\,, \nonumber\\
     {\cal B}_I &=& -2i\der_Ib\,, \nonumber\\
     {\cal C}_{IJ} &=& -\ft12 i(\phi
     +i\frac{b+\bar{b}}{\phi})\der_I\der_J(Xb)
     -\ft18 i\der_I\der_J(X^{-1}f)\,, \nonumber\\
     {\cal D}_I &=& {\rm Re}\bpl\ft13\phi^3\d_I\,^0
     +2i\phi X\der_I b
     -2\frac{b+\bar{b}}{\phi}\der_I(Xb)\bpr
     -\ft12{\rm Im}\,\der_I\bpl X^{-1}f\bpr \,.
\label{nonlin}\err
For the sake of clarity, we have absorbed the parameters
$\a$ and $\a_a$ into the functions $b$ and $g$ in the manner
described immediately after equation (\ref{zpdef}).
Substituting these functions into the Lagrangian (\ref{lagrangian}),
one may perform the duality transformation.  This is laborious,
but completely straightforward.  In this way we
obtain a dual description involving only vector multiplets,
which is characterized by the following holomorphic
prepotential,
\brr F &=& -\frac{1}{X^0}\bpl
    \ft13 X^1X^1X^1
    +\eta_{ab}X^1X^aX^b \bpr
    -\a X^1X^1
    +\a_a X^1X^a
    +f(X^0,X^a) \,.
\err
As discussed above, the quadratic terms
proportional to $\a$ and $\a_a$ represent
total derivatives, and the final term involves the background
self-interactions.  Notice that in this case the
prepotential is cubic in $X^1$. No higher-dimensional tensor theory 
is known that could possibly give rise to this coupling. 
\vspace{.3in}

\noindent{\it The linear vector-tensor multiplet:}\\[.05in]
As described previously, if $\eta_{11}=0$, implying the absense of the
$V\wedge {\rm d}V$ Chern-Simons coupling,
we obtain a vector-tensor multiplet which is
distinct from the nonlinear case just discussed.  In this case, it
is not possible to perform a field redefinition to remove all
of the $\eta_{1a}$ parameters.  Formally, this is indicated by the
presence of $\eta_{11}$ factors in denominators in (\ref{redef}).
In this case, the supersymmetric transformation
rules do not contain any terms more than linear in terms of the
vector-tensor component fields.
The functions
${\cal A}(\phi,X^I,\bar{X}^I)$,
${\cal B}_I(\phi,X^I,\bar{X}^I)$,
${\cal C}_{IJ}(\phi,X^I,\bar{X}^I)$, and
${\cal D}_I(\phi,X^I,\bar{X}^I)$
which define the linear multiplet and, more importantly,
the vector-tensor Lagrangian (\ref{lagrangian}) are now given by
the following expressions
\brr {\cal A} &=& -\ft12 g \,,\nonumber\\
     {\cal B}_I &=& -\frac{1}{g+\bar{g}}\bpl
     \phi\bar{g}\der_Ig
     -2i(b +\bar{b})\der_I g\bpr - 2i \der_I b\,, \nonumber\\
     \nonumber\\
     {\cal C}_{IJ} &=&
     -\frac{1}{g+\bar{g}}
     \bpl i\phi\bar{g}+2(b+\bar{b})\bpr\der_I\der_J(Xb)
     -\ft18 i\der_I\der_J(X^{-1}f)\,, \nonumber\\
     {\cal D}_I &=& \frac{1}{g+\bar{g}}{\rm Re}\bl
     \phi\bar{g}X\der_I(\phi g+4ib)
     -2i(b+\bar{b})\der_I[X(\phi g+4ib)]\br \,.
\err
As above, for the sake of clarity we have absorbed the parameters
$\a$ and $\a_a$ into the functions $b$ and $g$ in the manner
described immediately after equation (\ref{zpdef}).
Substituting these functions into the Lagrangian (\ref{lagrangian}),
and performing the duality transformation,
one obtains a dual description involving vector multiplets
alone, which is characterized by the following
prepotential,
\brr F &=& -\frac{1}{X^0}\bpl
    \ft12\eta_{1a}X^1X^1X^a
    +\eta_{ab}X^1X^aX^b \bpr
    -\a X^1X^1
    +\a_a X^1X^a
    +f(X^0,X^a)\,. \nonumber\\
\err
Again, as discussed above, the quadratic terms
involving $\a$ and $\a_a$ represent
total derivatives, while the last term involves the background
self-interactions.
Notice that in this case the prepotential has a term quadratic in
$X^1$, which cannot be suppressed. Such a term also arises from
the reduction of six-dimensional tensor multiplets to four dimensions.
In that case, the presence of the quadratic term is inevitable, because it
originates from the kinetic term of the tensor field. Although
there is no consistent action in six dimensions, because of the
self-duality of the tensors, this term can still be defined via
the various field equations \cite{FerMinSag}. Observe that we have
at least
three abelian vector fields coupling to the vector-tensor
multiplet, namely $W^0_\m$, $W^1_\m$ and $\eta_{1a}W^a_\m$.
\vspace{.1in}

The work presented in this paper represents an exhaustive
analysis of the $N=2$ vector-tensor multiplet with $8+8$
off-shell degrees of freedom, for the case of global
supersymmetry and local central charge,
in the presence of $n$ background vector
multiplets, one of which gauges the central charge.
We have presented the complete
and general transformation rules in this context, and
have shown that these actually include two distinct cases,
one of which is nonlinear in the vector-tensor components,
and the other of which is linear.
Furthermore we have constructed a supersymmetric action for
this system, and exhibited the bosonic part of this.  The
dual descriptions in terms of vector multiplets
have been obtained, and the respective prepotential
functions exhibited.  Neither in the nonlinear case nor in the
linear case do the associated K\"ahler spaces exhibit the factorization
property expected for the case of the heterotic string, at least in
the classical limit. As we discussed in the introduction, the
inability to construct certain couplings in off-shell formulations
with central charges, is not a new phenomenon and
is, at present, not understood. For the couplings constructed
in this paper, it is therefore not appropriate to associate the
vector-tensor  multiplet with the heterotic axion/dilaton complex.
However, as stressed in \cite{LSTY}, there can be additional
vector-tensor multiplets which are of nonperturbative origin.
Their contribution to the prepotential cannot be evaluated in
heterotic perturbation theory. They contribute in precisely two
distinct ways to the prepotential, corresponding to the two different
couplings found above. The linear vector-tensor multiplet coupling,
with terms quadratic and linear in the vector-tensor field $X^1$ is of
the type that one obtains from six-dimensional supergravity \cite{FerMinSag}.
Both vector-tensor couplings can be reproduced from the dual type-II
description, in addition to the perturbative heterotic couplings,
as was demonstrated in \cite{LSTY}. Note that our results are
applicable to any number of vector-tensor supermultiplets.

For the results presented above,
the fact that the central charge is realized locally enables the
inclusion of supergravity couplings in a rather
straightforward manner.
That program has also been carried out,
and the complete supergravity couplings will appear in a
seperate publication \cite{vtsg}.
\vspace{.1in}

We acknowledge useful discussions with S. Ferrara, J. Louis, D. L\"ust,
and A. Van Proeyen.  We especially thank B. Kleijn and R. Siebelink
for their help in the early stages of this work.
This work is supported by the European
Commission TMR programme ERBFMRX-CT96-0045. The work of M. Faux
is supported by FOM and also by the Institute for Theoretical
Physics at the University of Leuven. P.~Claus and P.~Termonia would like to
thank the Institute for Theoretical Physics at the University of Utrecht for
its warm hospitality.


\begin{thebibliography}{99}
\bibitem{sohnius} M.F. Sohnius, K.S. Stelle and P.C. West, Phys.
Lett. {\bf B92} (1980) 123.
\bibitem{DWKLL} B. de Wit, V. Kaplunovsky, J. Louis and D. L\"ust,
Nucl. Phys. {\bf B451} (1995) 53 (hep-th/9504006).
\bibitem{vt1} P. Claus, B. de Wit, M. Faux, B. Kleijn, R. Siebelink
and P. Termonia, Phys Lett. {\bf B373} (1996) 81
(hep-th/9512143).
\bibitem{LSTY} J. Louis, J. Sonnenschein, S. Theisen and S.
Yankielowicz,
Nucl. Phys. {\bf B480} (1996) 185
(hep-th/9606049).
\bibitem{FerMinSag} S. Ferrara, R. Minasian and A. Sagnotti,
Nucl.Phys. {\bf B474} (1996) 323 (hep-th/9604097).
\bibitem{DWLVP} B. de Wit, J.W. van Holten and A. Van Proeyen,
Nucl. Phys. {\bf B184} (1981) 77:\\ Errata {\bf B222} (1983) 516\\
B. de Wit, P.G. Lauwers and A. Van Proeyen,
Nucl. Phys. {\bf B255} (1985) 569.
\bibitem{HOW} A. Hindawi, B.A. Ovrut and D. Waldram, preprint UPR-712T,
(hep-th/9609016).
\bibitem{vtsg} P. Claus, B. de Wit, M. Faux, B. Kleijn, R. Siebelink
and P. Termonia, in preparation.
\bibitem{FVP} S. Ferrara and A. Van Proeyen,
Class. Quantum Grav. {\bf 6} (1989) L243.
\end{thebibliography}
\end{document}